\documentclass[aps,prl,showpacs,floats,twocolumn,floats,superscriptaddress,floatfix]{revtex4-1}
\usepackage{graphicx}
\usepackage{bm}
\usepackage{amsfonts}
\usepackage{color}
\usepackage{amsmath}    
\usepackage{epsfig}
\usepackage{subfigure}  
\usepackage{hyperref}   
\usepackage{bm}
\usepackage{amssymb}
\usepackage{soul,ulem}

\newcommand{\Wi}{\mbox{\textit{Wi}}}

\newcommand{\ictsaddress}{International Centre for
  Theoretical Sciences, Tata Institute of Fundamental Research,
  Bangalore 560089, India}
\newcommand{\UCA}{Universit\'e C\^ote d'Azur, CNRS, LJAD,
Nice 06100, France}
\newcommand{\losalomos}{Center for Nonlinear Studies, Los Alamos National Laboratory,
Los Alamos, NM 87545, U.S.A.}
\begin{document}
\title{Preferential Sampling of Elastic Chains in Turbulent Flows}
\author{Jason R. Picardo}
\email{jrpicardo@icts.res.in}
\affiliation{\ictsaddress}
\author{Dario Vincenzi}
\email{dario.vincenzi@unice.fr}
\affiliation{\UCA}
\author{Nairita Pal}
\email{nairitap2009@gmail.com}
\affiliation{\losalomos}
\author{Samriddhi Sankar Ray}
\email{samriddhisankarray@gmail.com}
\affiliation{\ictsaddress}
\begin{abstract}

A string of tracers, interacting elastically, in a turbulent
flow is shown to have a dramatically different behaviour when compared
to the non-interacting case. In particular, such an 
elastic chain shows strong preferential
sampling of the turbulent flow unlike the usual tracer limit:
an elastic chain is trapped in the vortical
regions. 
The degree of preferential sampling and its dependence on the elasticity of
the chain is quantified via the Okubo-Weiss parameter.
The effect of modifying the deformability of the chain, via the number of links that form it, is also examined.

\end{abstract}

\maketitle

The development of Lagrangian techniques, in experiments and theory, has
lead to major advances in our understanding of the complexity of turbulent
flows, especially at small scales~\cite{yeung,toschi,salazar}. What makes this possible is
the use of tracer particles 
which uniformly sample the flow and hence access the complete phase
space in which the dynamics resides. 
This feature of tracers depends, crucially, on the assumption that the
particles remain \textit{inertia-less} and \textit{point-like}. 
When some of these assumptions are relaxed, it may lead to
dissipative particle dynamics and preferential sampling of the
structures in a flow~\cite{pref-conc,jeremie-pof-jfm,celani,
monchaux,mehlig,inertial,collins-prefconc}.  
This is, for instance, the case for 
heavy, inertial particles, which 
show small-scale clustering and concentrate away from vortical regions.
Various phenomena
can influence the properties of inertial clustering in turbulence,
such as gravity \cite{ssr-gravity,collins-gravity}, turbophoresis \cite{turbo-1,turbo-2}, or the 
non-Newtonian nature of the fluid~\cite{polymer}.
Preferential sampling in turbulent flows may also emerge
as a result of the motility of particles, as in the case of 
gyrotactic~\cite{plankton}, interacting~\cite{flocking}, or
jumping~\cite{copepods} micro-swimmers.

We now propose a novel mechanism for preferential sampling
in turbulent flows which is induced by extensibility.
A simple model of an extensible object 
which retains enough internal
structure is a chain of tracers with an elastic coupling between
the nearest neighbours. We  show, remarkably,
that turning on such elastic interactions amongst tracers leads to very
different dynamics: unlike the case of 
non-interacting tracers, an elastic chain preferentially samples vortical regions
of the flow. We perform a systematic
study of this phenomenon and quantify, via the Okubo-Weiss parameter, the level
of preferential sampling and its dependence on the elasticity and deformability of the chain.

Harmonic chains have  been at the heart of several important problems in the
areas of equilibrium and non-equilibrium statistical physics. These have ranged
from problems in crystalline to amorphous transitions~\cite{amorphous},
electrical and thermal transport both in and
out-of-equilibrium~\cite{transport}, as well as understanding structural
properties of disordered and random systems~\cite{random}. Given the ubiquity
and usefulness of the elastic chain, it is surprising that the effect of a
turbulent medium on {\it long} chains has not been studied as extensively as
in other areas of non-equilibrium statistical physics. 

There is another reason why this study is important. The last decade or more
has seen tremendous advances in our understanding of heavy inertial particles
and their dynamics.  These were helped primarily by the pioneering results on
the issue of preferential concentration, which plays a dominant role in every
aspect of turbulent transport. In contrast, similar studies of
extended objects such as fibers are
recent~\cite{verhille,verhille-2,mazzino_fiber,jeremie_fiber}, even though they are just
as important and common place in nature and industry. However, the effect of
elasticity, which is intrinsic to extensible objects (just as inertia is to
finite-sized particles), in determining their dynamics in a turbulent flow is
an open question. In this Letter, we settle this question through a model which
is amenable to detailed numerical simulations.

\begin{figure*}
\includegraphics[width=1\textwidth]{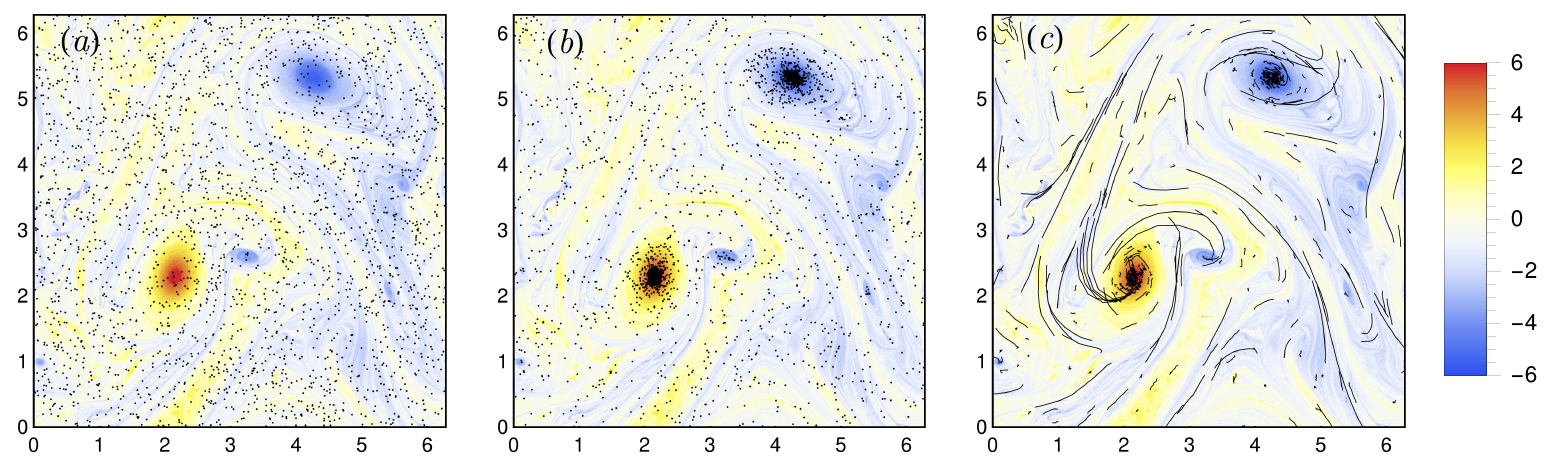}
\caption{Representative snapshots showing the positions of a subset of
chains ($N_{\rm L} = 9$ and $L_{\rm m} = 4$) overlaid on the vorticity field of the carrier 
turbulent flow (with $\ell_f = 2$ and $t_f = 2.3$). Panels (a) and (b) show the centers of mass of the chains for $\Wi = 0.04$ and $\Wi = 0.9$ respectively. Panel (c) shows entire chains for $\Wi = 0.9$. Parameter values: $\nu = 10^{-6}$, $\mu = 10^{-2}$, and $F_0 = 0.2$.} 
\label{fig:snapCM}
\end{figure*}

We generalize a well-studied model for polymeric chains, 
the Rouse model \cite{bird,doi,ottinger}, which consists
of a sequence of 
$N_{\rm b}$ identical beads connected through (phantom) elastic links
with its nearest left and right
neighbours; the two end beads are free. Starting from Newton's equation for a 
single bead and incorporating the effect of the fluid Stokesian drag, elastic interactions, and 
thermal noise, the dynamics is most conveniently expressed in terms of
the center of mass of the chain $\bm X_c=(\bm x_1+\dots+\bm
x_{N_{\mathrm{b}}})/N_{\rm b}$ ($\bm x_1,\dots,\bm x_{N_{\mathrm{b}}}$ denote
the positions of the beads) and
the separation vector $\bm r_j=\bm x_{j+1}-\bm x_j$ ($j=1,\dots,N_{\rm b}-1$)
between the $j$-th and $(j+1)$-th bead.
For arbitrary $r_j$,
the general form of the equations of motion of such a chain  
in a velocity field $\bm u(\bm x,t)$, in the absence of inertia,  are:
\begin{subequations}
\begin{align}
\dot{\bm X}_c&=\dfrac{1}{N_{\rm b}} \left[ \sum_{i=1}^{N_{\rm b}} \bm u(\bm x_i,t)
+\sqrt{\dfrac{r_0^2}{2\tau}}\sum_{i=1}^{N_{\rm b}}\bm\xi_i(t) \right] \label{eq:cm}\\
\dot{\bm r}_j&=\bm u(\bm x_{j+1},t)-\bm u(\bm x_j,t) +\sqrt{\dfrac{r_0^2}{2\tau}}\,[\bm \xi_{j+1}(t)-\bm \xi_j(t)] \nonumber \\
&\qquad -\dfrac{1}{4\tau}(2f_j\bm r_j-f_{j+1}\bm r_{j+1}-f_{j-1}\bm r_{j-1}) \label{eq:disp}
\end{align}
\label{eq:motion}
\end{subequations}%
where
$f_j=({1-\vert\bm r_j\vert^2/r_{\rm m}^2})^{-1}$ are the 
standard FENE (Finitely-Extensible-Nonlinear-Elastic) interactions \cite{bird},
which are linear for small separations and diverge quadratically for larger ${r}_j$ to ensure that 
inter-bead separations remain bounded by the maximum length $r_{\rm m}$.
Thus, at any given instant in time, the length of the chain $R \equiv
\sum_{j}^{N_{\rm L}}r_j < L_{\rm m}$, where
$L_{\rm m} =N_{\rm L}r_{\rm m}$ is the maximum or contour chain length 
($N_{\rm L} = N_{\rm b} - 1$ denotes the number of links).
The nonlinear elastic links are further
characterised by their relaxation time $\tau$,
which, in turn, determines the effective relaxation time $\tau_{\rm chain} = (N_{\rm b}+1)N_{\rm b}\tau/6$ of the chains~\cite{collins}. 
The thermal fluctuations on each bead
are modelled by independent white noises $\bm\xi_i(t)$ ($i=1,\dots,N_{\rm b}$),
whose amplitude $r_0$ sets the equilibrium length of the chain in the absence
of flow.  Although the effect of thermal fluctuations on the motion of the
center of mass is negligible in a turbulent flow, their effect on the
separation vectors is essential: thermal fluctuations prevent the chain from
collapsing into a point-like particle and hence a tracer.  Finally, inter-bead
hydrodynamic interactions are ignored.

\begin{figure*}
\includegraphics[width=1.0\textwidth]{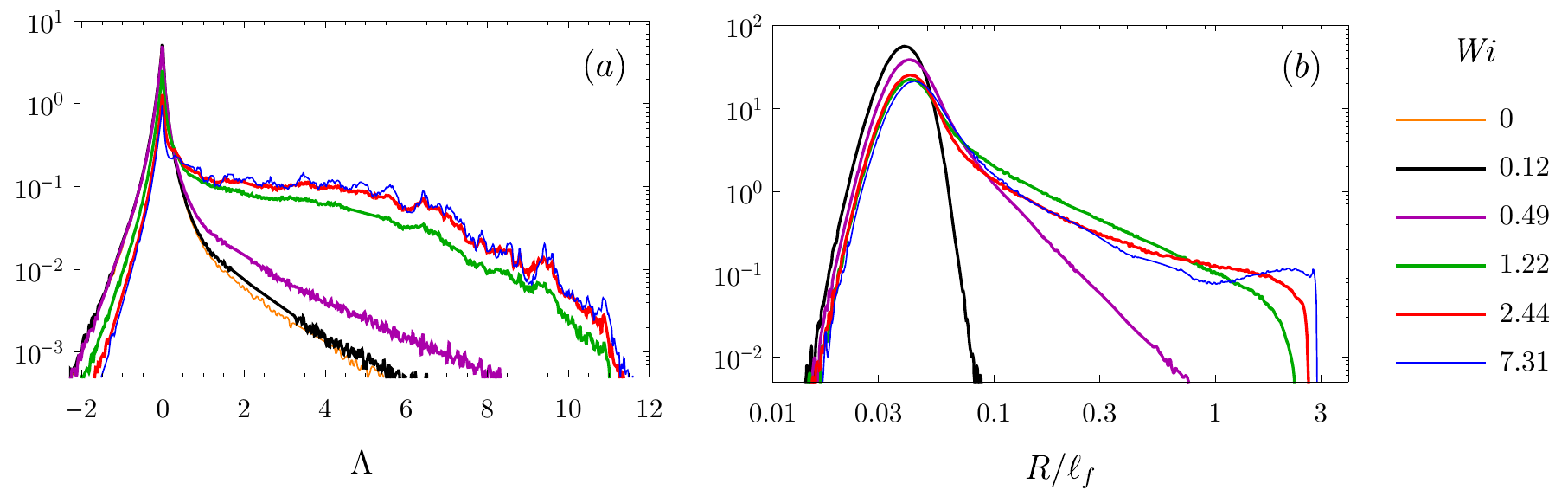}
\caption{Lagrangian pdfs of (a) the
Okubo-Weiss parameter $\Lambda$ and (b) the 
scaled lengths of the chains $R/\ell_f$ for different values of $\Wi$.
Parameter values:  $\ell_f = 1.25, t_f = 1.35, L_m = 3.75, N_L = 9$, $\nu = 10^{-6}$, $\mu = 10^{-2}$, and $F_0 = 0.2$.}
\label{fig:pdf}
\end{figure*}

We note that such a generalised model ensures that when the
separations between the beads are small enough for the velocity differences to
be approximated as $\bm u(\bm x_{j+1},t)-\bm u(\bm x_j,t) \approx \mathcal{A}\cdot\bm r_j$, where $\mathcal{A}_{ik}=\partial_k u_i$ is the velocity gradient tensor, then Eqs.~\eqref{eq:motion} reduce to the Rouse
model with FENE links smaller than the viscous scale at all
times. This model is commonly used in simulations of polymer chains in turbulent flows~\cite{zhou,collins,watanabe,gupta,massah}. 
By considering the full velocity difference between adjacent beads,
Eqs.~\eqref{eq:motion} describe an
elastic chain that may extend into the inertial range of turbulence or even
beyond the integral scale (see Refs.~\cite{mazzino,piva,schumacher,av16} for an
analogous generalization of the dumbbell ($N_{\rm b}=2$) model).  

What is the effect of a turbulent velocity field $\bm u$
on the motion of such chains?
To answer this, we solve, by using a pseudo-spectral method with a 2/3 de-aliasing rule, 
the two-dimensional Navier-Stokes equation on a square grid with $1024^2$ collocation points and $2\pi$ periodic 
boundary conditions. 
We drive the system to a homogeneous and isotropic, turbulent, 
statistically steady state through an external, deterministic
force $f=-F_0 k_{f} \cos(k_{f} x)$ ($F_0$ is the amplitude and $k_{f}$ the
energy-injection scale in Fourier space, which  sets the typical size of the
vortices $\ell_f=2 \pi k_f^{-1}$).
Forcing at small wavenumbers ensures that the vortices are fairly
large: this allows us 
to clearly illustrate the issues of preferential sampling which are
central to this work.
We use a small Ekmann-friction coefficient $\mu$ (in addition to a coefficient of kinematic viscosity 
$\nu$) to prevent
pile-up of energy at the large scales due to inverse cascade.
Consequently the turbulent flow is in the direct-cascade regime.
The definition of $\ell_f$ also allows us to characterise the
stretching ability of the flow
in terms of the dimension-less Weissenberg number $\Wi
= \tau_{\rm chain}/t_f$, where  $t_f=\ell_f/\sqrt{2E}$ is the turnover time
scale of the large vortices ($E$ is the mean kinetic energy of the
flow).  

The temporal evolution of the chain (Eqs.~\eqref{eq:motion})
is done by a second-order Runge-Kutta scheme,
augmented by a rejection algorithm~\cite{ottinger}
to avoid numerical instabilities due to the divergence
of the nonlinear force for $\vert\bm r_j\vert$ approaching $r_{\rm m}$.
A bilinear scheme is used to interpolate, 
from the Eulerian grid, the fluid velocity at the 
typically off-grid positions of the beads~\cite{PrasadPRL,JamesSR}. 
In order to observe the regime of preferential sampling,
we choose parameters for the chain which ensure that its equilibrium
length in a quiescent flow is similar to the enstrophy dissipation
scale, while $\ell_f < L_{\rm
m} < 2\pi$.

This model of an elastic chain in a turbulent flow is the ideal
setting, theoretically and numerically, to investigate the natural
interplay between the relative importance of Lagrangian (uniform) mixing and the
elasticity of the links. It is this competing effect that leads to a
surprising preferential sampling of the flow by the chain, hitherto not
observed. 

It is important to stress here that we chose two-dimensional
turbulent flows to take advantage of their long-lived vortical structures,
which help to convincingly illustrate this new phenomenon of preferential
sampling. We have checked in several simulations that this phenomenon persists even in three-dimensional
turbulence, becoming increasingly prominent as the Reynolds number is raised and intense vortex filaments proliferate. However, as in the case of preferential concentration of inertial
particles, the effect is most convincingly brought out in two-dimensional
flows.

To illustrate this phenomenon, we begin by 
randomly seeding $5\times 10^4$ chains into the flow
and study their evolution in time for different $\mathit{Wi}$. 
(We evolve a large number of chains simultaneously for the purpose of 
visualizing their sampling behaviour and for
obtaining good statistics of the chain dynamics; we do not
describe the collective motion of an ensemble of chains,
which would interact with each other hydrodynamically or
by direct contact.)
In Fig.~\ref{fig:snapCM} we show the center-of-mass positions 
at an instant of time overlaid on the vorticity field of the turbulent
flow for $\Wi = 0.04$ and $\Wi = 0.9$.  It is immediately apparent
that for the case of small elasticity
the chains behave like tracers and distribute evenly 
(Fig.~\ref{fig:snapCM}(a)). However, for larger $\Wi$
there is a preferential sampling of the
vortical regions (Fig.~\ref{fig:snapCM}(b)).
 
Figure~\ref{fig:snapCM}(c) demonstrates the
coupling between the translational and the extensional dynamics of the chain
by showing
a snapshot in which the entire chains, and not
just the centers of mass, are overlaid on the vorticity field. This figure
emphasises the strong correlation between the positions of elongated
chains with regions of low vorticity, where the straining flow stretches out the chains.
In contrast, the chains that encounter
vortices tend to curl up and contract to a much smaller size. These 
strikingly different phenomena are best seen in a video of the time
evolution of the chains~\cite{movieChain}. All
of this suggests the following picture: a stretched chain is
more likely to leave straining zones and coil up in vortical regions. 

The above observations can be quantified via a Lagrangian
approach by measuring the statistics of the extension $R$ and the Okubo-Weiss
parameter $\Lambda$
along the trajectories of the centers of mass of the
chains.  We recall that, for incompressible flows,  $\Lambda \equiv \rm{det} \mathcal{A}/\langle \omega^2
\rangle = (\omega^2-\sigma^2)/4 \langle \omega^2
\rangle$ (here rescaled by the mean enstrophy $ \langle \omega^2
\rangle$), where ${\bm \omega} = \nabla \times {\bm u}$ is the vorticity and $\sigma$ is the strain rate, given by $\sigma^2 = 2\mathcal{S}_{ij}\mathcal{S}_{ij}$ where $\mathcal{S}= (\mathcal{A}+\mathcal{A}^{\rm T})/2$. The sign of $\Lambda$ uniquely determines
the local flow geometry: For positive $\Lambda$, the flow is vortical; for
negative $\Lambda$ it is extensional~\cite{OW,PrasadPRL}.  Values of $\Lambda$
near zero correspond to either a quiescent or  a
shearing flow, which has comparable amounts of vorticity and straining.

In Fig.~\ref{fig:pdf} we show the Lagrangian probability distribution
function (pdf) of (a) $\Lambda$ and (b) 
$R$ for different values of $\Wi$.  The $\Lambda$ distribution for
tracers ($\Wi =0$) is also shown for comparison (its positive skewness is a consequence
of the strongest velocity gradients occurring in intense
vortical zones~\cite{prasad,anupam,dhruba}).  We find strong quantitative
evidence that increasing elasticity leads to a chain preferentially sampling
vortical regions of the flow (Fig.~\ref{fig:pdf}(a)).  This is accompanied by
an increase in the probability of highly stretched configurations 
(Fig.~\ref{fig:pdf}(b)).

\begin{figure}
\includegraphics[width=1.0\columnwidth]{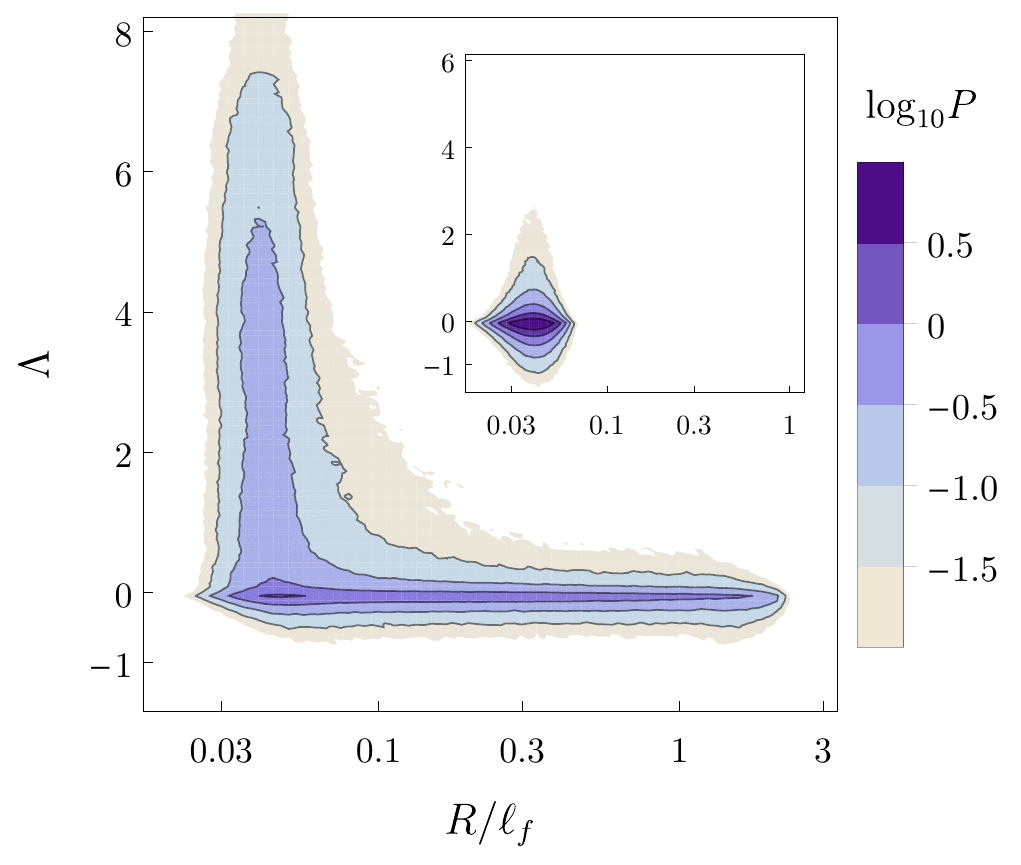}
\caption{Joint pdf of $\Lambda$ sampled by the centers of mass
of the chains and $R/\ell_f$ for $\Wi = 1.22$ and (inset) $\Wi = 0.12$.}
\label{fig:joint_pdf}
\end{figure}

The key to understanding the phenomenon of preferential
sampling lies in the correlation between the translational
and the extensional dynamics of a chain.
This is  quantified through the joint pdf
$P(R,\Lambda)$,
which shows that when its center of mass is
in vortical regions, a chain is in a
contracted state (Fig.~\ref{fig:joint_pdf}).  
The velocity terms in Eq.~\eqref{eq:cm} can therefore be Taylor-expanded
about $\bm X_c(t)$, and Eq.~\eqref{eq:cm} reduces to
the equation of motion
of a tracer. Thence, the center of mass follows the flow and remains
trapped in the vortex. Contrastingly, 
away from vortical regions, i.e., in straining or
shearing regions, an extensible chain is highly stretched (Fig.~\ref{fig:joint_pdf}). 
As time proceeds, a large-$\Wi$ chain
stretches to lengths so long that eventually it is unable to follow
the rapidly evolving straining zones. On departing from these zones,
such a chain is likely to encounter a vortex and 
begin to coil up (seen clearly in the video~\cite{movieChain}). The links of the chain
that enter the vortex shrink and follow the rotational flow,
eventually leading to the entrapment of the entire chain within the vortex.
A stiff chain (small-$\Wi$), which remains short in straining zones, samples the negative values of $\Lambda$ more,  
leading to annular contours (Fig.~\ref{fig:joint_pdf}, inset).

The stretching out of a chain in straining zones may thus be seen as a
precursor to its entrapment inside vortices. This explains
why strong preferential
sampling of vortices occurs only when $\Wi$ is so large that there is a significant
probability for chains to be stretched beyond $\ell_f$ (see
Fig.~\ref{fig:pdf} for $\Wi = 0.49,1.22$).  
On the other hand, if the maximum length 
$L_m < \ell_f$, then one would expect this mechanism to fail and the chain 
to uniformly sample the flow for all $\Wi$. 
We have confirmed this hypothesis in our simulations but do not show the results 
for brevity.

The results, so far, suggest a complete picture for the
dynamics of an elastic chain in a turbulent flow: A chain with a sufficient
degree of elasticity---defined as the ratio of elastic and fluid 
time scales---preferentially samples the flow. 
But is this effect truly independent of the characteristic
\textit{length} scales present in the system? The short answer is no, and the
ability of a chain to preferentially sample the flow is determined by the
relative magnitudes of its typical inter-bead separation (which for a
fixed value of $L_{\rm m}$ depends only on $N_{\rm b}$) and the
characteristic fluid length scale $\ell_f$. 

For large values of $\Wi$, as we have seen, 
the typical inter-bead separation approaches $L_{m}/N_{L}$. 
For the results reported so far,  
$N_{\rm b}$ was such that $L_{m}/N_{L} \ll \ell_f$. As $N_{\rm b}$
decreases, however, we will eventually obtain  characteristic inter-bead
separations of the order of $L_{m}/N_{L} > \ell_f$. In this setting,
typically, no two neighboring beads will be able to reside in a vortex
simultaneously. Hence the mechanism for preferential sampling 
will fail, and the chain will start sampling the flow uniformly
once more. This suggests that, apart from the role of elasticity, the
dynamics of a chain for large values of $\Wi$ ought to depend on a second
dimensionless number $\Phi \equiv {L_{\rm m}}/({N_{\rm
L} \ell_f)}$, such that for $\Phi > 1$ there should be uniform sampling,
while for $\Phi < 1$ preferential sampling.

\begin{figure}
\includegraphics[width=1.00\columnwidth]{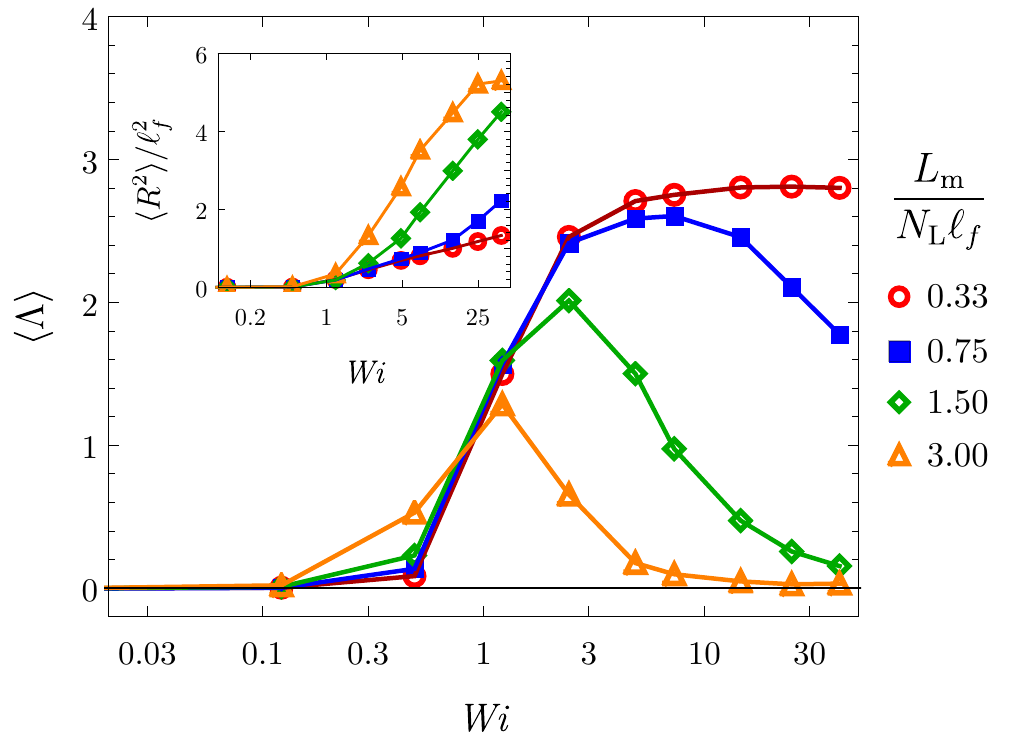}
	\caption{$\langle\Lambda\rangle$ vs $\Wi$ for different values of
	$\Phi \equiv {L_{\rm m}}/({N_{\rm L} \ell_f})$; we use $N_{\rm L}$ = 9, 4, 2, and 1 (dumbbell), 
	while keeping $L_{\rm m} = 3.75$ and $\ell_f = 1.25$ fixed.
	Inset: The variance
	of $R/\ell_f$ vs $\Wi$. The flow parameters are the same as those in Fig.~\ref{fig:pdf}.}
\label{fig:avgNlinks}
\end{figure}

Direct evidence for this is presented in Fig.~\ref{fig:avgNlinks}, which shows
the mean value of the Okubo-Weiss parameter, $\langle \Lambda \rangle$, as a
function of $\Wi$, for different values of $\Phi$.  For smaller $\Phi$, there
is an increase in $\langle\Lambda \rangle$ that eventually saturates for $\Wi
\gtrsim 5$.  However, for larger values of $\Phi$, $\langle \Lambda \rangle$
increases initially~\cite{footnote}---indicating preferential sampling of the vortical regions
of the flow---before decreasing again to reflect uniform sampling (see
video~\cite{movieDb}). The chain, however, continues to stretch as the
elasticity increases for all values of $\Phi$, and its mean square extension
does not show any non-monotonic behavior (inset of Fig.~\ref{fig:avgNlinks}).
At large $\Wi$ a chain is indeed typically longer for larger values of $\Phi$
owing to the reduced level of preferential sampling of vortices and
correspondingly lower probability of being in a contracted state.

Figure~\ref{fig:avgNlinks} lends itself to an intuitively appealing picture
of the motion of a chain. For a given turbulent flow (characterised by
$t_f$ and $\ell_f$), whether or not a chain of a given elasticity and maximum length
$L_{\rm m}$ may coil---and hence preferentially sample---depends on
the number of links which form the chain. In particular, for a highly extended
chain (large $\Wi$), decreasing the number of links limits the ability of the
chain to coil up into vortices.
This result highlights the importance of {\it deformability} of extensible
chains. In our study, we have considered a freely-jointed chain.
However,
the bending stiffness of the chain may be modeled by introducing a
potential force that is a function of the internal angle between neighboring
links and restores them to an anti-parallel configuration~\cite{bird}.  Such
stiffness would prevent the
chain from coiling up and, therefore, would
reduce preferential sampling in a manner analogous to that 
of increasing $\Phi$
in our freely-jointed model.  

The elastic chain is a simplified model for various physical systems, which include,
{\it inter alia}, fibers, microtubules, and algae in marine environments. Such systems of course
present additional properties that were not taken into account here, such as
the inertia and
the stiffness of the system, hydrodynamic and excluded-volume
interactions between different portions of it, and
the modification of the flow generated by the motion of the system.
These effects will certainly change quantitative details of the dynamics, but
the mechanism at the origin of preferential sampling identified here is of general validity.
It indeed relies on a few basic ingredients: the system must be extensible, its equilibrium size should
be smaller than $\ell_f$ and its contour length greater than $\ell_f$.

Finally, we cannot avoid mentioning that the preferential concentration of 
inertial particles through dissipative dynamics is completely different from the mechanism that 
we report here. Hence it is tempting to investigate the interplay between the competing effects of inertia and elasticity
in future studies of inertial extensible objects in turbulent flows.

\begin{acknowledgments}
We thank Abhishek Dhar, Anupam Kundu, and Rahul Pandit for useful suggestions and discussions.
We acknowledge the support of the Indo-French Centre for Applied Mathematics
(IFCAM).  SSR acknowledges DST (India) project
ECR/2015/000361 for financial support. JRP acknowledges travel support
from the ICTS Infosys Collaboration grant. The simulations were
performed on the ICTS clusters {\it Mowgli} and {\it Mario}, as well as
the work stations from the project ECR/2015/000361: {\it Goopy} and
{\it Bagha}.
\end{acknowledgments}



\begin{thebibliography}{0}%
\makeatletter
\providecommand \@ifxundefined [1]{%
 \@ifx{#1\undefined}
}%
\providecommand \@ifnum [1]{%
 \ifnum #1\expandafter \@firstoftwo
 \else \expandafter \@secondoftwo
 \fi
}%
\providecommand \@ifx [1]{%
 \ifx #1\expandafter \@firstoftwo
 \else \expandafter \@secondoftwo
 \fi
}%
\providecommand \natexlab [1]{#1}%
\providecommand \enquote  [1]{``#1''}%
\providecommand \bibnamefont  [1]{#1}%
\providecommand \bibfnamefont [1]{#1}%
\providecommand \citenamefont [1]{#1}%
\providecommand \href@noop [0]{\@secondoftwo}%
\providecommand \href [0]{\begingroup \@sanitize@url \@href}%
\providecommand \@href[1]{\@@startlink{#1}\@@href}%
\providecommand \@@href[1]{\endgroup#1\@@endlink}%
\providecommand \@sanitize@url [0]{\catcode `\\12\catcode `\$12\catcode
  `\&12\catcode `\#12\catcode `\^12\catcode `\_12\catcode `\%12\relax}%
\providecommand \@@startlink[1]{}%
\providecommand \@@endlink[0]{}%
\providecommand \url  [0]{\begingroup\@sanitize@url \@url }%
\providecommand \@url [1]{\endgroup\@href {#1}{\urlprefix }}%
\providecommand \urlprefix  [0]{URL }%
\providecommand \Eprint [0]{\href }%
\providecommand \doibase [0]{http://dx.doi.org/}%
\providecommand \selectlanguage [0]{\@gobble}%
\providecommand \bibinfo  [0]{\@secondoftwo}%
\providecommand \bibfield  [0]{\@secondoftwo}%
\providecommand \translation [1]{[#1]}%
\providecommand \BibitemOpen [0]{}%
\providecommand \bibitemStop [0]{}%
\providecommand \bibitemNoStop [0]{.\EOS\space}%
\providecommand \EOS [0]{\spacefactor3000\relax}%
\providecommand \BibitemShut  [1]{\csname bibitem#1\endcsname}%
\let\auto@bib@innerbib\@empty
\end{thebibliography}%


\begin{thebibliography}{10}

\bibitem{yeung} P. K. Yeung, Annu. Rev. Fluid Mech. {\bf 34}, 115 (2002).

\bibitem{toschi}
F. Toschi and E. Bodenschatz, Annu. Rev. Fluid Mech. {\bf 41}, 375 (2009).

\bibitem{salazar}
J. P. L. C. Salazar and L. R. Collins, Annu. Rev. Fluid Mech. {\bf 41},
405 (2009).

\bibitem{pref-conc} E. K. Longmire and J. K. Eaton, J. Fluid Mech. {\bf 236}, 217 (1992)

\bibitem{jeremie-pof-jfm}
J. Bec, Phys. Fluids {\bf 15}, 81 (2003);
J. Fluid Mech. {\bf 528}, 255 (2005).

\bibitem{celani}
J. Bec. A. Celani, M. Cencini, and S. Musacchio,
Phys. Fluids {\bf 17}, 073301 (2005).

\bibitem{collins-prefconc}
J. Chun, D. L. Koch, S. L. Rani, A. Ahluwalia, and L. R. Collins, 
J. Fluid Mech. {\bf 536}, 219 (2005).

\bibitem{inertial} 
J. Bec, L. Biferale, M. Cencini, A. Lanotte, S. Musacchio, and F. Toschi, Phys. Rev. Lett. {\bf 98}, 084502 (2007). 

\bibitem{monchaux}
R. Monchaux, M. Bourgoin, and A. Cartellier, Int. J. Multiphase Flow {\bf 40}, 1 (2012).

\bibitem{mehlig}
K. Gustavsson and B. Mehlig, Adv. Phys. {\bf 65}, 1 (2016).

\bibitem{ssr-gravity}
J. Bec, H. Homann, and S. S. Ray,
Phys. Rev. Lett. {\bf 112}, 184501 (2014).

\bibitem{collins-gravity}
G. H. Good, P. J Ireland, G. P. Bewley, E. Bodenschatz, L. R. Collins, and Z. Warhaft, 
J. Fluid Mech. {\bf 759}, R3 (2014).

\bibitem{turbo-1}
S. Belan, I. Fouxon, and G. Falkovich, Phys. Rev. Lett. {\bf 112}, 234502 (2014).

\bibitem{turbo-2}
F. De Lillo, M. Cencini, S. Musacchio, and G. Boffetta
Phys. Fluids {\bf 28}, 035104 (2016).

\bibitem{polymer}
F. De Lillo, G. Boffetta, S. Musacchio,
Phys. Rev. E {\bf 85}, 036308 (2012).

\bibitem{plankton}
W. M. Durham, E. Climent, M. Barry, F. De Lillo, G. Boffetta, M. Cencini, and R. Stocker,
Nature Communications {\bf 4}, 2148 (2013).

\bibitem{flocking}
  A. Choudhary, D. Venkataraman, and S. S. Ray, 
    Europhys. Lett. {\bf 112}, 24005 (2015).

\bibitem{copepods}
H. Ardeshiri, I. Benkeddad, F. G. Schmitt, S. Souissi, F. Toschi and E. Calzavarini, 
Phys. Rev. E {\bf 93}, 043117 (2016).

\bibitem{amorphous} 
Z. H. Stachurski, Materials {\bf 4}, 1564 (2011);
B. Illing, S. Fritschi, H. Kaiser, C. L. Klix, G. Maret, and P. Keim Proc. Nat. Acad. Sci. {\bf 114}, 1856 (2017).

\bibitem{transport}
	A Dhar, Phys. Rev. Lett. {\bf 86}, 5882 (2001); 
	S. Lepri, R. Livi, A. Politi, Phys. Rep. {\bf 377}, 1 (2003);
	A Dhar, Adv. Phys. {\bf 57}, 457 (2008); 
	A. Kundu, S. Sabhapandit, and A. Dhar, J. Stat. Mech., P03007 (2011).	

\bibitem{random} A. M. Mayes, Macromolecules {\bf 27}, 3114 (1994); K. Binder, J. Baschnagel, and W. Paul, Prog. Polymer Sc. {\bf 28}, 115 (2003);
		P. Chaudhuri, S. Karmakar, C. Dasgupta, H. R. Krishnamurthy, and A. K. Sood, Phys. Rev. Lett., {\bf 95}, 248301 (2005); 
		T. Sridhar, D. A. Nguyen, R. Prabhakar, and J. R. Prakash, Phys. Rev. Lett. {\bf 98}, 167801 (2007).

\bibitem{verhille} 
C. Brouzet, G. Verhille, and P. Le Gal, 
Phys. Rev. Lett. {\bf 112}, 074501 (2014).

\bibitem{verhille-2}
G. Verhille and A. Bartoli, Exp. Fluids {\bf 57}, 117 (2016).

\bibitem{mazzino_fiber}
M. E. Rosti, A. A. Banaei, L. Brandt, and A. Mazzino,
Phys. Rev. Lett. {\bf 121}, 044501 (2018).

\bibitem{jeremie_fiber}
S. Allende, C. Henry, and J. Bec, arXiv:1805.05731



\bibitem{bird}
R. B. Bird, C. F. Curtiss, R. C. Armstrong, and O. Hassager,
{\it Dynamics of Polymeric Liquids}
(John Wiley and Sons, New
York, 1977), Vol. 2

\bibitem{doi} M. Doi  and S. F. Edwards, {\it The theory of polymer dynamics} (Oxford University Press,
1986).

\bibitem{ottinger}
 H. C. \"{O}ttinger,
{\it Stochastic Processes in Polymeric Fluids}
(Springer, Berlin, Germany, 1996).

\bibitem{massah}
H. Massah, K. Kontomaris, W. R. Schowalter, and T. J. Hanratty,
Phys. Fluids A {\bf 5}, 881 (1993)

\bibitem{zhou}
Q. Zhou and R. Akhavan, J. Non-Newtonian Fluid Mech. {\bf 109}, 115 (2003).

\bibitem{gupta}
V. K. Gupta, R. Sureshkumar, and B. Khomami,
Phys. Fluids {\bf 16}, 1546 (2004).

\bibitem{collins} 
J. Jin and L. R. Collins, New J. Phys. {\bf 9}, 360 (2007).

\bibitem{watanabe}
T. Watanabe and T. Gotoh, Phys. Rev. E {\bf 81}, 066301 (2010).

\bibitem{mazzino}
M. De Lucia, A. Mazzino, and A. Vulpiani,
Europhys. Lett. {\bf 60}, 181 (2002).

\bibitem{piva}
M. F. Piva and S. Gabanelli, J. Phys. A: Math. Gen. {\bf 36},
4291 (2003).

\bibitem{schumacher}
J. Davoudi and J. Schumacher, Phys. Fluids {\bf 18},
025103 (2006).

\bibitem{av16}
A. Ahmad and D. Vincenzi, Phys. Rev. E {\bf 93}, 052605 (2016).




\bibitem{PrasadPRL} P. Perlekar, S. S. Ray, D. Mitra, and R. Pandit, Phys. Rev. Lett. {\bf 106} 054501 (2011).


\bibitem{JamesSR} M. James and S. S. Ray, Sci. Rep. {\bf 7}, 12231 (2017)

\bibitem{movieChain} 
\url{https://youtu.be/etLuK6ovAqk}
		
\bibitem{OW} A. Okubo, Deep-Sea Res. Oceanogr. Abstr. {\bf 17}, 445
	(1970); J. Weiss, Physica (Amsterdam) {\bf 48D}, 273 (1991).

\bibitem{prasad}
P. Perlekar and R. Pandit, New J. Phys. {\bf 11}, 073003 (2011)

\bibitem{anupam}
A. Gupta, P. Perlekar, and R. Pandit,
Phys. Rev. E {\bf 91}, 033013 (2015)

\bibitem{dhruba}
D. Mitra and P. Perlekar,
Phys. Rev. Fluids {\bf 3}, 044303 (2018)

\bibitem{footnote}
See also Refs.~\cite{mazzino,piva} for dumbbells in cellular flows.

\bibitem{movieDb} 
Online movie showing uniform sampling of highly stretched dumbbells: \url{https://youtu.be/YdFXMFFU1qU}

\end{thebibliography}
\end{document}